# Strong Photothermoelectric Response and Contact Reactivity of the Dirac Semimetal ZrTe$_5$

*François Léonard\*[†], Wenlong Yu[‡], Kimberlee C. Collins[†], Douglas L. Medlin[†], Joshua D. Sugar[†], A. Alec Talin[†], Wei Pan[‡]*

[†]Sandia National Laboratories, Livermore, CA, 94551, United States

[‡]Sandia National Laboratories, Albuquerque, NM, 87185, United States

\*fleonar@sandia.gov

**Abstract**

The family of three-dimensional topological insulators opens new avenues to discover novel photophysics and to develop novel types of photodetectors. ZrTe$_5$ has been shown to be a Dirac semimetal possessing unique topological electronic and optical properties. Here we present spatially-resolved photocurrent measurements on devices made of nanoplatelets of ZrTe$_5$, demonstrating the photothermoelectric origin of the photoresponse. Due to the high electrical conductivity and good Seebeck coefficient, we obtain noise-equivalent powers as low as 42 pW/Hz$^{1/2}$ at room temperature for visible light illumination at zero bias. We also show that these devices suffer from significant ambient reactivity such as the formation of a Te-rich surface region driven by Zr oxidation, as well as severe reactions with the metal contacts. This reactivity results in significant stresses in the devices, leading to unusual geometries that are useful for gaining insight into the photocurrent mechanisms. Our results indicate that both the large photothermoelectric response and reactivity must be considered when designing or interpreting photocurrent measurements in these systems.

**Introduction**

Novel topological phases of matter are attracting much attention recently[1] and are opening new avenues for basic science discovery and novel approaches for future devices[2, 3]. Within the family of topological materials, Dirac semimetals[4-7] have shown intriguing properties such as the chiral magnetic effect[8], pressure-induced superconductivity[9], and suppression of electron backscattering[10]. They are also of special interest because they can serve as precursors for other topological phases such as Weyl semimetals and topological superconductors. A number of materials have been identified as Dirac semimetals including $Cd_3As_2$, $Ni_3Bi$, and $ZrTe_5$.

While the electronic transport properties of topological materials have been studied[11, 12], there are only a few experimental[13-17] and theoretical[18, 19] studies of their optoelectronic properties. The theoretical studies have considered the role of spin-momentum locking in topological insulators for generating helically-dependent photocurrent and have suggested that Weyl semimetals could show a large photoresponse due to symmetry breaking and their unusual electronic bandstructure[19]. Experimentally, most of the optoelectronic work has focused on topological insulators such as $Bi_2Te_3$, with the exception of a very recent study[20] on $Cd_3As_2$. In the case of $ZrTe_5$, there has been previous work on the temperature-dependent electronic transport[21] and thermoelectric[22] properties, but to date, there is no report of photocurrent measurements in this Dirac semimetal. Establishing the basic photocurrent response is important because extracting novel behavior based on topological physics requires a careful consideration of all the mechanisms that can lead to photocurrent.

In addition to its interest for topological physics, $ZrTe_5$ is also of interest as a photodetector material. Indeed, several recent studies have shown that the photothermoelectric effect can be quite strong in thin film and layered materials[20, 23, 24] so we may expect that $ZrTe_5$, with its good thermoelectric properties may be promising for realizing high performance room temperature photodetectors. However, to our knowledge such photodetectors have not been previously studied.

In this manuscript, we present experimental results of the visible photocurrent response of $ZrTe_5$ nanoplatelets using scanning photocurrent microscopy (SPCM). At zero bias we observe strong photoresponse when the light is focused on the metal contacts, with a sign (and reversal) at the source and drain contacts consistent with the photothermoelectric effect. We engineer the thermal environment to achieve a room-temperature Noise-Equivalent Power (NEP) as low as 42 pW/Hz$^{1/2}$. Furthermore, we show that the strong interaction of Zr with oxygen and interaction of the Pd electrical contact material with the $ZrTe_5$ leads to significant morphological and compositional changes in the material and the devices, causing significant stresses that can deform devices. Such deformations turn out to be useful to further demonstrate the photothermoelectric origin of the photocurrent. Our results are important for the future design of devices based on $ZrTe_5$ because they show that material interactions critically determine performance and reliability. Furthermore, our results establish that devices based on $ZrTe_5$ are excellent photodetectors when the photothermoelectric effect is harnessed.

Figures 1a,b show a schematic of the devices considered in this work, and an optical image of a particular device. Nanoplatelets were created by mechanically exfoliating $ZrTe_5$ crystals[25], and were then deposited on Si/SiO$_2$ substrates. Electron beam lithography was used to fabricate several Pd electrodes (300nm thickness) on top of the nanoplatelets, allowing for different measurement combinations between pairs of electrodes in order to gain insight into the role of channel resistance and electrode spacing on the photoresponse. Channel lengths were between ~ 1 micron and a few tens of microns while the platelets were between 95nm and 700nm in thickness. After fabrication and during all measurements the devices were kept in ambient. No heat treatment was applied to the devices. Figure 1c shows the current-voltage characteristic for the device of Fig. 1b, measured at room temperature in ambient between the two end electrodes. The resistance of only 400 Ohms points to the highly conductive nature of this material.

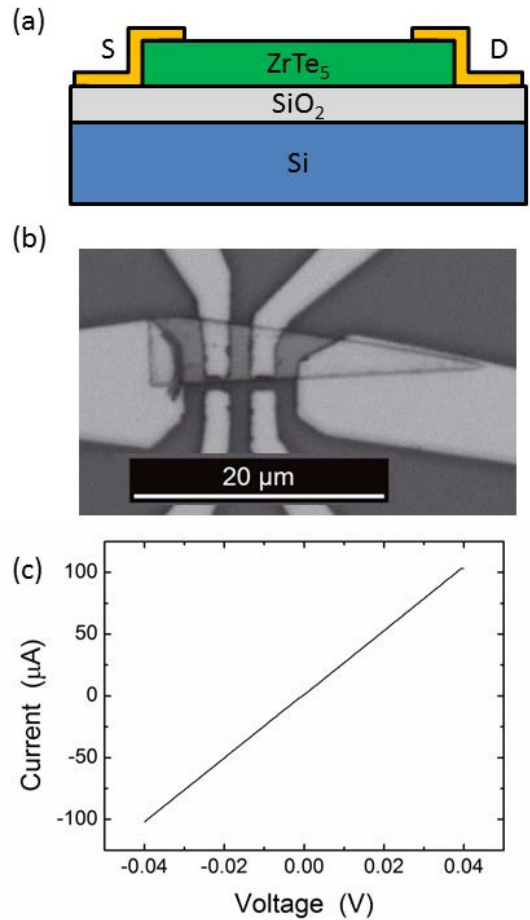

**Figure 1.** (a) Cross-sectional schematic of a ZrTe$_5$ nanoplatelet on Si/SiO$_2$ contacted by source and drain electrodes. (b) Optical image of a fabricated device, using Pd for the metal contacts. (c) Current versus voltage measured at room temperature in ambient.

To explore the optoelectronic response of these devices, we used scanning photocurrent microscopy (SPCM) whereby a laser is focused to the diffraction limit and rastered over the device as the photocurrent and reflection image are recorded as a function of the laser position. Details of the approach can be found in previous publications[26, 27]. The acquisition of the reflection image concomitantly with the photocurrent map allows the unambiguous alignment of the photocurrent map with the structural features of the devices, such as electrode and nanoplatelet edges. Measurements were performed with a CW red laser (632 nm) in ambient.

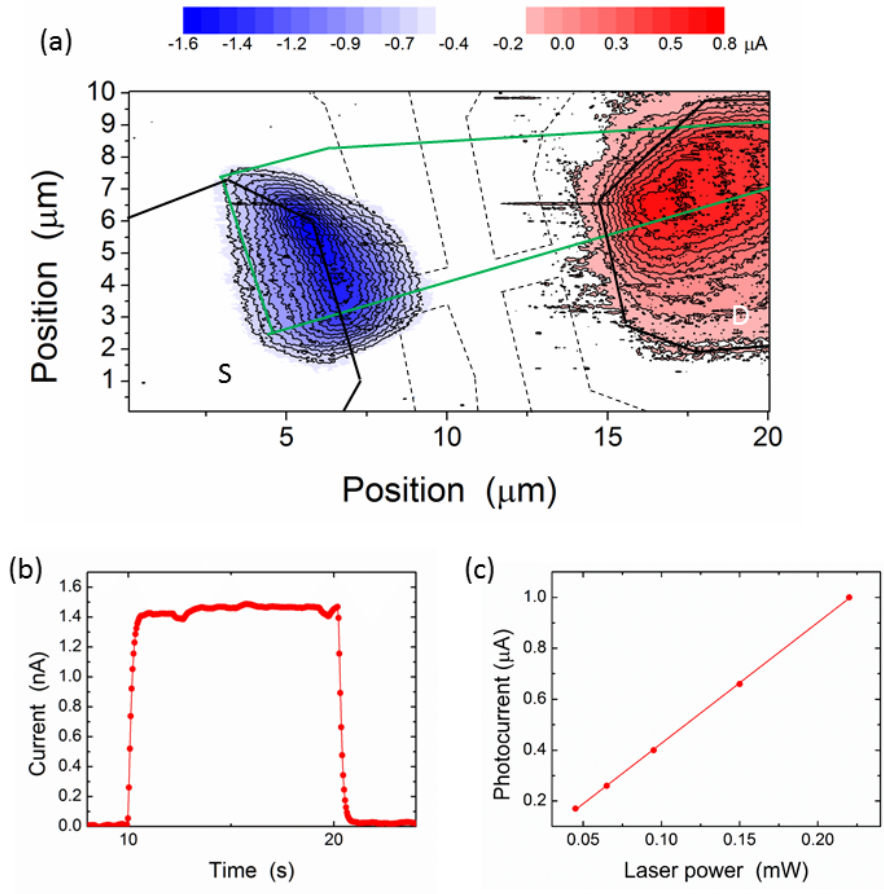

**Figure 2.** (a) Scanning photocurrent microscopy of a ZrTe$_5$ nanoplatelet on Si/SiO$_2$ obtained with a red laser focused to the diffraction limit. The nanoplatelet edges are shown with green lines. The solid black lines denote the source and drain electrode edges. The dashed black lines are the edges of electrically floating electrodes. (b) Time dependence of the photocurrent when the laser is focused at the peak photocurrent on the right side. Rise and fall times are determined by the 300 ms pre-amp integration time. (c) Power dependence of the photocurrent when the laser is focused at the peak photocurrent on the right side.

Figure 2 shows the SPCM map of the device of Fig. 1b obtained at zero source-drain bias. In this measurement the source and drain electrodes are the two electrodes at the ends of the nanoplatelet, outlined with solid black lines in the figure. The other four electrodes (dashed black lines in the figure) were left floating. The SPCM map clearly shows two peaks of opposite sign located where the nanoplatelet intersects the source and drain contacts. We probed the time dependence of the signal by focusing the laser at the positive peak and turning the laser ON and OFF. Figure 2b shows an example of the resulting time response when the pre-amp integration time is 300 ms. Additional measurements with smaller pre-amp integration times allow us to extract a device response time of 5 ms (see Supporting Information for details). Furthermore, we find a linear dependence of the photocurrent on the laser power (Fig. 2c). The linear I-V characteristics, the SPCM map combined with the relatively slow time scale, and the linear dependence on power imply that the photothermoelectric effect is the dominant mechanism[26]. (Further evidence based on the unique device structures will be provided below). We also note that heating of the contact leading to increased thermal excitation of carriers over a Schottky barrier cannot explain the results because we observe a photoresponse at zero bias; indeed, heating of a Schottky barrier does not generate current at zero bias. Photoemission from the metal to the semiconductor can also be ruled out because the metal is too thick (300nm) to allow light to penetrate efficiently to the vertical metal-semiconductor junction. And photocurrent generation due to band-bending near the contact can also be eliminated since photocurrent is observed when the laser is over the metal electrodes.

The photothermoelectric effect arises when the laser causes local heating and a temperature difference between the two electrodes, leading to a voltage $\Delta V = -S\Delta T$; this can be more specifically written as[26] $\Delta V_{meas} = V_S - V_D = (S_{NP} - S_M)(T_D - T_S) \approx S_{NP}(T_D - T_S)$ where $\Delta V_{meas}$ is the voltage measured by the external measurement circuit, $S_{NP}$ is the nanoplatelet Seebeck coefficient, $S_M \ll S_{NP}$ is the metal Seebeck coefficient, $T_D$ is the drain temperature, and $T_S$ is the source temperature. The Seebeck coefficient of as-grown crystalline ZrTe$_5$ has been reported to be +125 µV/K at room temperature[22]. i.e. it is usually a p-type semiconductor at room temperature. The sign of our photocurrent data agrees with this positive Seebeck coefficient. Since the nanoplatelets should be thick enough to prevent finite size effects and because of their high quality as demonstrated below, we expect the Seebeck coefficient to be similar to that of high-quality bulk material. Using the value of the bulk Seebeck coefficient we extract from Fig. 1c and Fig. 2c a temperature increase of 14.4 K/mW. This is in good agreement with previous measurements of laser heating of metal contacts in nanodevices, where simultaneous measurements of the Seebeck coefficient and the photoresponse allowed the extraction of the temperature increase[26]. (SPCM measurements on additional devices and electrode configurations are presented in the Supporting Information.)

Because of the large Seebeck coefficient and relatively high temperature increase, the magnitude of the photocurrent is quite large in this device. For example, from Fig. 2c we obtain a responsivity of 4.5 mA/W (or 1.8 V/W using the device resistance). This, combined with the low resistance leads to excellent photodetector performance. A typical measure of photodetector performance is the Noise-Equivalent-Power (NEP) defined as the ratio of the dark noise spectrum to the responsivity. Since the devices function at zero bias, the dark noise spectrum is simply the thermal noise spectrum given by $\sqrt{4k_BTR}$ where $T$ is temperature and $R$ the device resistance. Thus, the large responsivity combined with the low resistance leads to a room-temperature Noise-Equivalent-Power (NEP) of 1.2 nW/Hz$^{1/2}$. (For a detailed discussion of the figure of merit of photothermoelectric detectors, see Ref. [28]. This reference and references therein also show experimentally that for zero-bias devices the noise is limited by the thermal noise.)

For a photothermoelectric mechanism, we expect that reducing the heat transfer to the environment would lower the NEP at the expense of a longer response time. We tested this idea by fabricating devices over trenches in Si substrates. Figure 3a shows an optical micrograph of one such device with the trench appearing in green. An SEM image shows that not only is the middle portion of the ZrTe$_5$ platelet suspended, but so is the end near the large contact (right side of SEM image). This suspension of the material leads to an impressive photoresponse: as shown by the SPCM map of Fig. 3c, the maximum response near the suspended electrode gives a photocurrent of tens of microamps, more than an order of magnitude larger than the device shown in Fig. 1b at the same laser power. This photoresponse translates into a large responsivity of 0.13 A/W (or 74 V/W using the device resistance) which gives a room-temperature NEP of 42 pW/Hz$^{1/2}$. Furthermore, the trench suspension leads to a photoresponse in the channel that extends more than 10 microns away from the left contact. Additional evidence for the photothermoelectric behavior comes from the temporal behavior (Fig. 3c) with a response time of about 0.5 second, much slower than the non-suspended devices, as a consequence of the reduced heat dissipation to the environment.

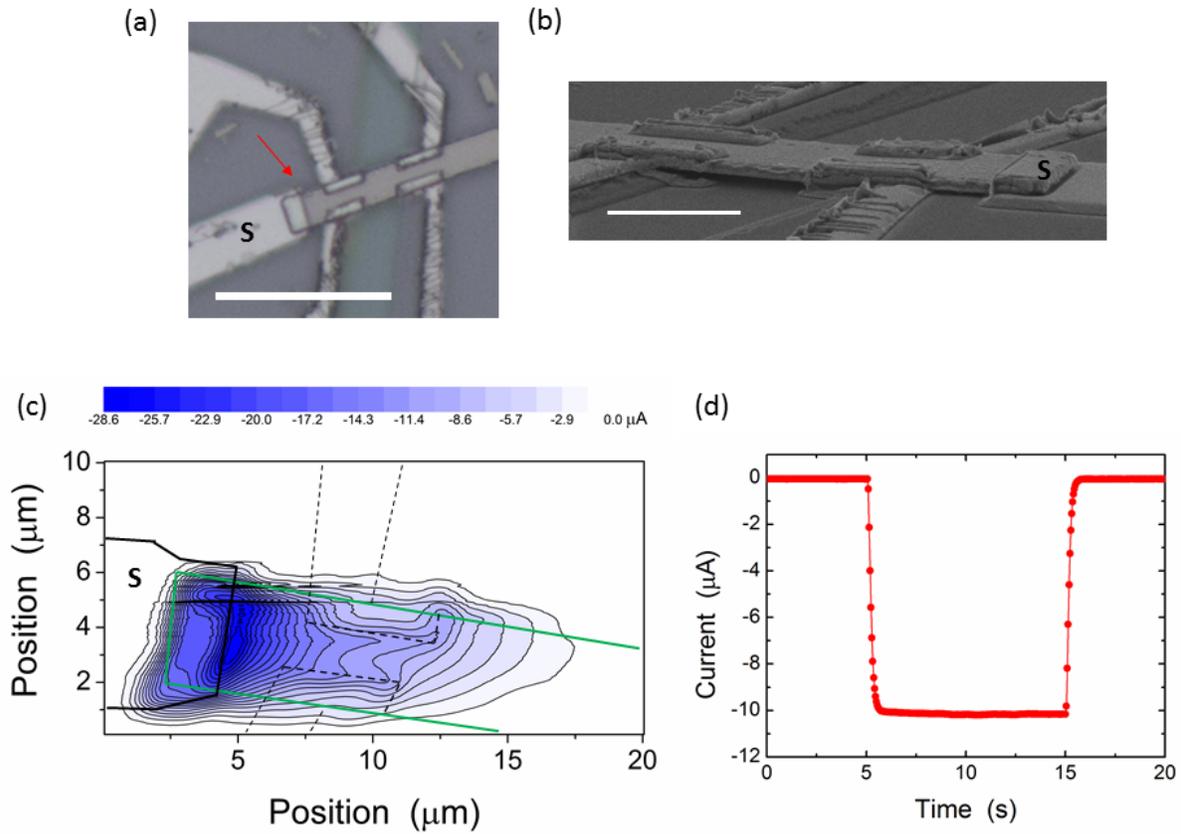

**Figure 3.** (a) Optical micrograph of a ZrTe$_5$ nanoplatelet partially suspended over a trench in the Si substrate. Scale bar: 20 microns. Only the source side of the device is shown due to the long overall length of the device. (b) SEM image of the device, taken from the direction indicated by the red arrow in panel (a). Scale bar: 5 microns. (c) Scanning photocurrent microscopy of the device obtained with a red laser (power = 0.22 mW) focused to the diffraction limit. The nanoplatelet edges are shown with green lines. The solid black line denotes the source electrode edges. The dashed black lines are the edges of floating

electrodes. The mapped area is around the large source electrode near the red arrow in panel (a). (d) Time dependence of the photocurrent when the laser is focused on the source electrode.

Our investigations of the optoelectronic response revealed unusual device structural properties that turn out to dominate the photoresponse. Indeed, while a dependence of the photoresponse on the nanoplatelet thickness was expected based on heat dissipation arguments, such a correlation was not observed due to the strong device deformations. For example, Fig. 4a shows a SPCM map for a different device obtained at zero bias. We observe a photocurrent several microns away from the edge of the left contact, both in the channel and over the contact region. This behavior is unexpected since this device was not suspended. This unusual behavior of the photoresponse originates from a dramatic deformation of the device. Figure 4b shows SEM images of the device taken after the SPCM experiments, which were conducted after the device was stored in ambient for several days. The images show a strong texture on the surface of the $ZrTe_5$ as well as on the Pd metal in contact with the $ZrTe_5$. This texture is also accompanied by severe deformations of the device: for example, some of the metal contacts have been almost destroyed, and part of the $ZrTe_5$ nanoplatelet has curled and lifted from the substrate even where metal was deposited on top. Because the nanoplatelet is lifted from the substrate the thermal dissipation to the environment is reduced, allowing the local laser heating to spread to longer distances. In a sense the lifted end of the nanoplatelet serves as an optical antenna that funnels the generated heat to what remains of the nanoplatelet/metal contact. The observation of photoresponse tens of microns away from the edge of the metal/$ZrTe_5$ contact where the nanoplatelet is no longer in contact with the substrate clearly supports the photothermoelectric mechanism. Note that the stress in the device should be located mainly where the metal has been deposited on top of the nanoplatelets. Therefore, we expect the Seebeck coefficient to remain unchanged in the channel. Since the relevant Seebeck coefficient is the spatially-averaged Seebeck over the device length, variations of the Seebeck coefficient due to stress are not expected to significantly affect the effective Seebeck coefficient.

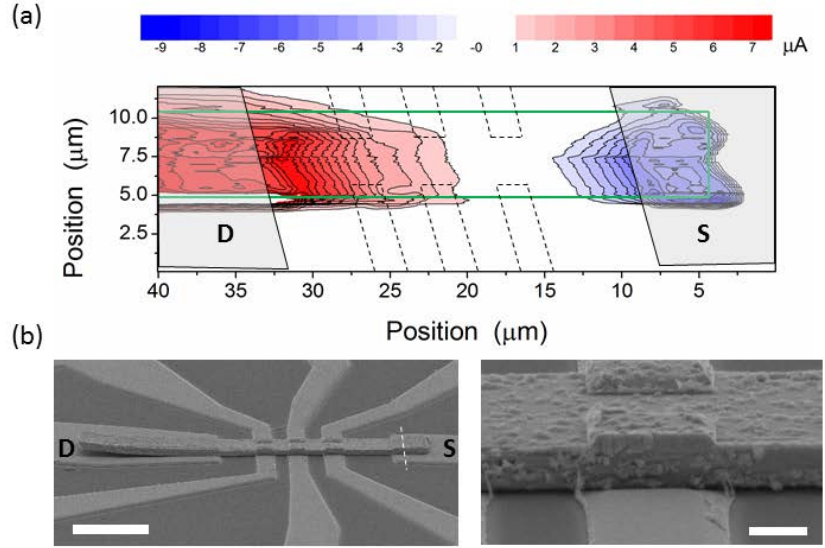

**Figure 4.** (a) Scanning photocurrent microscopy of a ZrTe$_5$ nanoplatelet on Si/SiO$_2$ obtained with a focused red laser. The nanoplatelet edges are shown with green lines. The solid black lines denote the source and drain electrode edges which are also shaded in grey. The dashed black lines are the edges of floating electrodes. (b) Low and high magnification images of the device showing the impact of reactivity on device structure. The white dashed line on the left panel of (b) shows the position of TEM cross-section through the Pd contact discussed later in the text. Scale bars: 10 microns left panel, 1 micron right panel.

To gain insight into the device reactivity we performed Raman mapping of the device of Fig. 4 using a 532 nm laser. Figure 5a shows an optical image of the device including the focused Raman laser while Fig. 5b shows the Raman spectrum acquired when the laser was located over the ZrTe$_5$ between electrodes as indicated with the star in Fig. 5a. The Raman spectrum shows several features that can be compared with known Raman lines of ZrTe$_5$[29], Te[30], and ZrO[31]. The main lines for ZrTe$_5$ are found at 37 cm$^{-1}$ and 182 cm$^{-1}$ as indicated with the taller blue lines in Fig. 5b. While ZrTe$_5$ also has lines between these two values, they are much smaller in intensity. Therefore, the strong peaks observed at ~110 cm$^{-1}$ and 120 cm$^{-1}$ arise from Raman modes of other species. In particular, Te has prominent lines at 120 cm$^{-1}$ and 140 cm$^{-1}$. Further evidence for the presence of Te comes from the line at 92 cm$^{-1}$ which can be clearly seen in the inset of Fig. 5b. Additionally, we find lines that are specific to ZrO, notably the line at 103 cm$^{-1}$ and a broad peak that is close to the line at 265 cm$^{-1}$. Thus, the Raman characterization suggests that ZrTe$_5$ reacts with oxygen to form ZrO, leaving free Te that coalesces to form a surface phase. The distribution of Te appears to be uniform on the surface of the nanoplatelet: the Raman maps in Fig. 5c,b show how the line at 123 cm$^{-1}$ is found throughout the surface of the ZrTe$_5$, but not elsewhere on the electrodes or the Si substrate. We observed device reactivity in all of the devices that we fabricated, and several additional examples are presented in the Supporting Information.

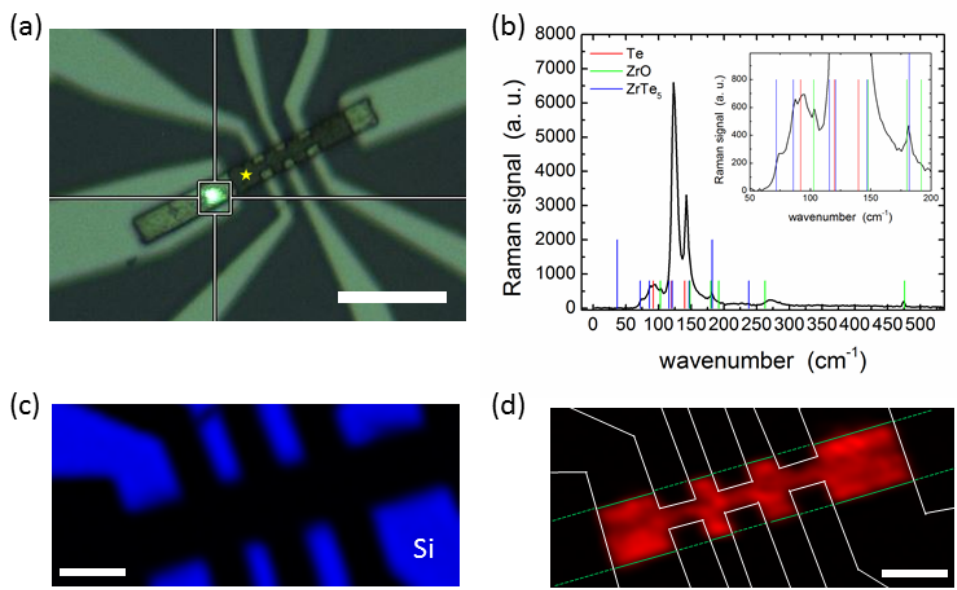

**Figure 5.** (a) Optical micrograph of a ZrTe$_5$ device for Raman measurements. The green laser is visible in the crosshair. Scale bar: 20 microns. (b) Raman spectrum acquired with the laser focused on the ZrTe$_5$ between contacts, indicated with the yellow star in panel (a). Vertical lines indicate the positions of Raman lines from ZrTe$_5$, ZrO, and Te. Prominent lines for ZrTe$_5$ are indicated with taller vertical lines. (c) Raman map obtained from the intensity of the Si peak. Scale bar: 5 microns. (d) Raman map obtained from the intensity of the peak at 123 cm$^{-1}$. Scale bar 5 microns.

To further elucidate the nature of the reaction mechanisms, we used scanning transmission electron microscopy (STEM) to investigate the structure of a ZrTe$_5$ platelet and its interface with the Pd contact. A cross-sectional TEM specimen was prepared by focused ion beam (FIB) milling through the Pd/ZrTe$_5$ interface of the right-most contact of the device shown in Figure 4b. Figure 6 shows a HAADF-STEM image (a) and a corresponding energy dispersive x-ray spectroscopy (EDS) map (b) of the device in cross-section and projected along the direction of the channel. The interface between the Pd and the $ZrTe_5$ platelet is quite rough; the gap between the two layers may reflect partial delamination due to reactivity or during the TEM specimen preparation. Selected area electron diffraction (SAED) (Figure 6c) shows the rings from the polycrystalline, randomly oriented Pd metal and a single-crystal pattern from the underlying $ZrTe_5$ platelet. Both the single-crystal pattern and the high-resolution STEM image (Fig. 6d) are consistent with the expected[32] orthorhombic structure of $ZrTe_5$ and show that the channel direction is oriented along a [1 0 0] direction (i.e., the *a*-axis) and that the surface normal of the platelet is along the [0 1 0] direction (i.e., the *b*-axis). The SAED pattern also exhibits a weak ring corresponding to a d-spacing of 0.29 nm. While this diffraction ring is insufficient on its own to identify the species, in consideration of the Raman results discussed above, we note that {111} reflections[33, 34] for the cubic phases of ZrO and $ZrO_2$ correspond to d-spacings of 0.27 nm and 0.29 nm, respectively. Finally, the EDS results (Fig. 6b) also show a strong Pd signal in the upper portion of the $ZrTe_5$. For instance, analysis of EDS spectra collected from regions of the platelet above the crack marked in Fig. 6a, give Pd concentrations as high as 28 at. % (details are provided in the Supporting Information). This result indicates significant diffusion of Pd from the electrical contact into the platelet. As illustrated in Fig. 6d, the $ZrTe_5$ crystal structure possesses large, open interstices between the individual $ZrTe_5$ units. This open structure may provide the crystal the capability of accommodating large concentrations of solute atoms. More generally, our measurements suggest that $ZrTe_5$ is a highly reactive material, leading to modifications of its free surface and significant modifications of the electrical contacts.

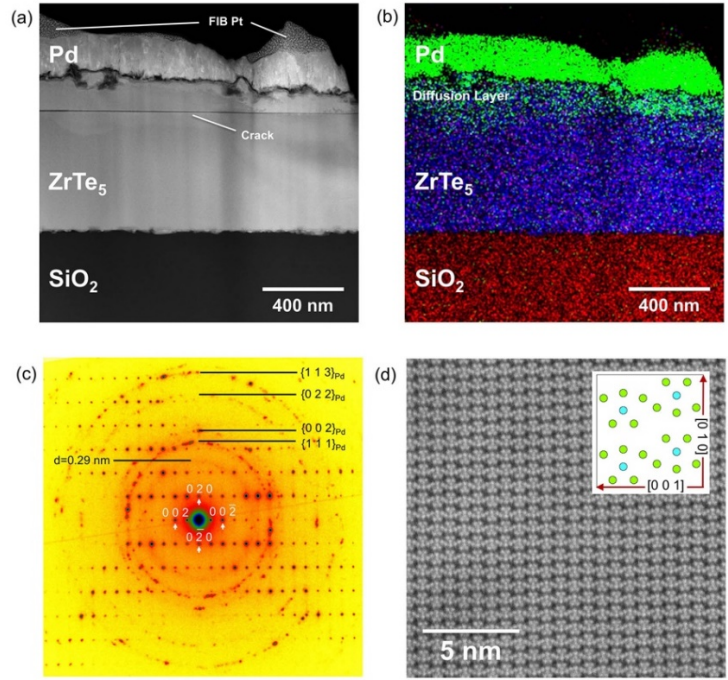

**Figure 6.** (a) HAADF-STEM image of interface between Pt contact and ZrTe$_5$ platelet from region of device indicated in Fig. 4(b). (b) EDS map from the same region as (a) showing intensity of signals from Pd-L$\alpha$ (Green), Zr-K$\alpha$ (magenta), Te-L$\alpha$ (blue), and Si-K$\alpha$ (red) x-rays. (c) Selected Area Electron diffraction pattern from the ZrTe$_5$ platelet and Pd contact layer. The single crystal pattern, indexed with white characters, is consistent with the [1 0 0] zone of ZrTe$_5$. The ring pattern is consistent with polycrystalline Pd. An additional ring, with d=0.29 nm is also detected. (d) High resolution HAADF-STEM image taken from a region in lower half of the ZrTe$_5$ platelet. The inset to (d) shows a schematic of the ZrTe$_5$ crystal structure projected along the [1 0 0] zone (Green=Te, Light Blue=Zr).

**Conclusion**

In conclusion, we find that ZrTe$_5$ nanoplatelet devices show a very strong photothermoelectric response under visible light illumination at zero bias. Surprisingly, without detailed engineering of the device configuration or the materials, we measure a responsivity as high as 74 V/W and a NEP as low as 42 pW/Hz$^{1/2}$ at room temperature. This can be compared with recent measurements on other thin film and layered materials where the photothermoelectric effect at contacts is the dominant photoresponse mechanism. For example[23, 24], carbon nanotube thin films and graphene have shown responsivities up to 10 V/W and NEP of 1 nW/Hz$^{1/2}$. A relevant comparison is with $Cd_3As_2$ nanoplatelets, where scanning photocurrent microscopy data implies responsivities on the order of 0.1 V/W and NEP of about 3 nW/Hz$^{1/2}$. In addition, since the optical absorption occurs in the contact material, we expect the photoresponse to be broadband. Thus, $ZrTe_5$ is a promising material to realize high performance, room-temperature, broadband, photodetectors.

The observation of such a large photothermoelectric response in $ZrTe_5$ implies that future experiments exploring exotic physics in these systems will need to carefully mitigate this effect. Similarly, the strong reactivity of the material is another important factor that emerges from our work, emphasizing that future work will have to carefully control the history, measurement environment, and contact metallurgy for $ZrTe_5$ samples. To ensure reliability of electrical contacts to $ZrTe_5$, it may be important to employ a barrier layer to prevent interdiffusion, or to carefully study metal interactions to ensure device reliability. Finally, detailed studies of the temporal evolution of the metal reactivity and the impact on device performance would provide additional insight on the degradation mechanisms and their potential mitigation.

## ASSOCIATED CONTENT

**Supporting Information**. Additional examples of photocurrent measurements, device reactivity, and photodetector performance evaluation. Data of time response measurements. Details of FIB sample preparation and TEM analysis. This material is available free of charge via the Internet at http://pubs.acs.org.

## AUTHOR INFORMATION

**Corresponding Authors**

*Email: fleonar@sandia.gov

**Notes**

The authors declare no competing financial interest.

ACKNOWLEDGEMENTS

We thank Warren York for preparation of the FIB specimens. Work supported by the Laboratory Directed Research and Development program at Sandia National Laboratories. WY and WP were partially supported by the U.S. Department of Energy, Office of Science, Basic Energy Sciences, Materials Sciences and Engineering Division. This work was performed, in part, at the Center for Integrated Nanotechnologies, an Office of Science User Facility operated for the U.S. Department of Energy (DOE) Office of Science. Sandia National Laboratories is a multimission laboratory managed and operated by National Technology and Engineering Solutions of Sandia, LLC., a wholly owned subsidiary of Honeywell International, Inc., for the U.S. Department of Energy's National Nuclear Security Administration under contract DE-NA-0003525.

**(Table of Contents figure)**

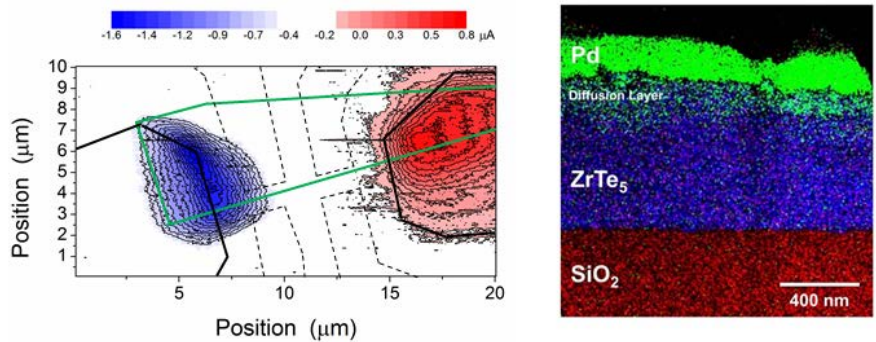

# Supporting Information:

# Strong Photothermoelectric Response and Contact Reactivity of the Dirac Semimetal ZrTe$_5$

*François Léonard\*[†], Wenlong Yu[‡], Kimberlee C. Collins[†], Douglas L. Medlin[†], Joshua D. Sugar[†], A. Alec Talin[†], Wei Pan[‡]*

[†]Sandia National Laboratories, Livermore, CA, 94551, United States

[‡]Sandia National Laboratories, Albuquerque, NM, 87185, United States

[\*fleonar@sandia.gov](*fleonar@sandia.gov)

*1. Additional SPCM images and NEP results*

In this section we present additional SPCM images taken on several different devices and with different electrode configurations, as well as the measured responsivity, resistance, and Noise-Equivalent-Power. We note the presence of apparent photocurrent outside of the devices in some of the SPCM maps; this originates from a drift of the dark current during the map acquisition time.

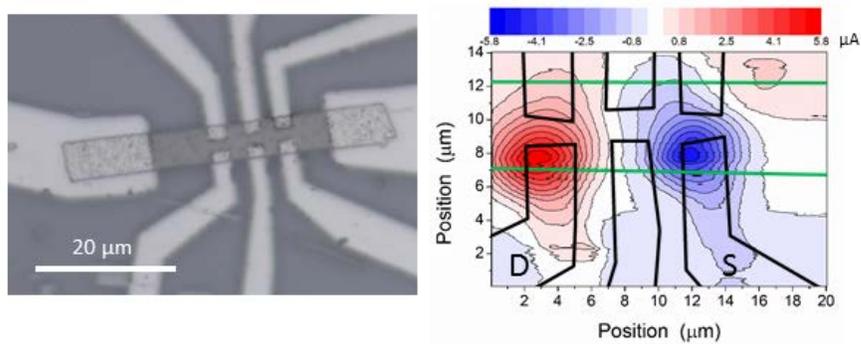

Figure S1: Optical image of a device (Dev SI-1) and the resulting SPCM map. The electrodes between which the photocurrent is measured are indicated with the labels "S" for source and "D" for drain.

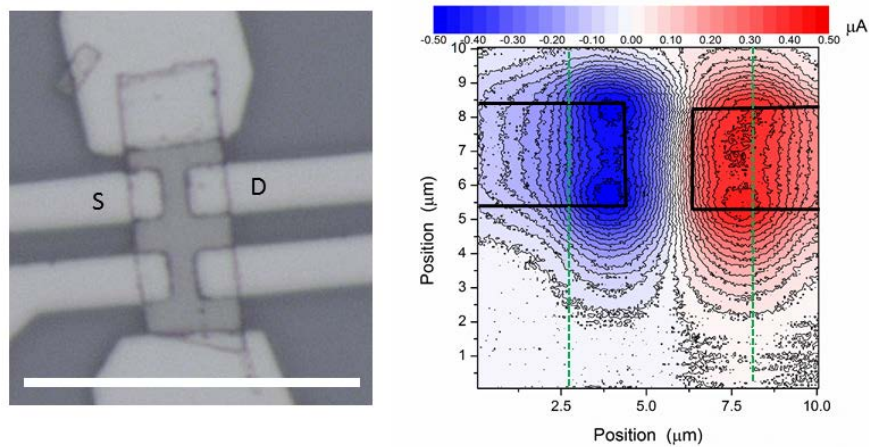

Figure S2: Optical image of a device (Dev SI-2) and the resulting SPCM map.
Scale bar in the optical image is 20 microns. The electrodes between which the photocurrent is measured are indicated with the labels "S" for source and "D" for drain.

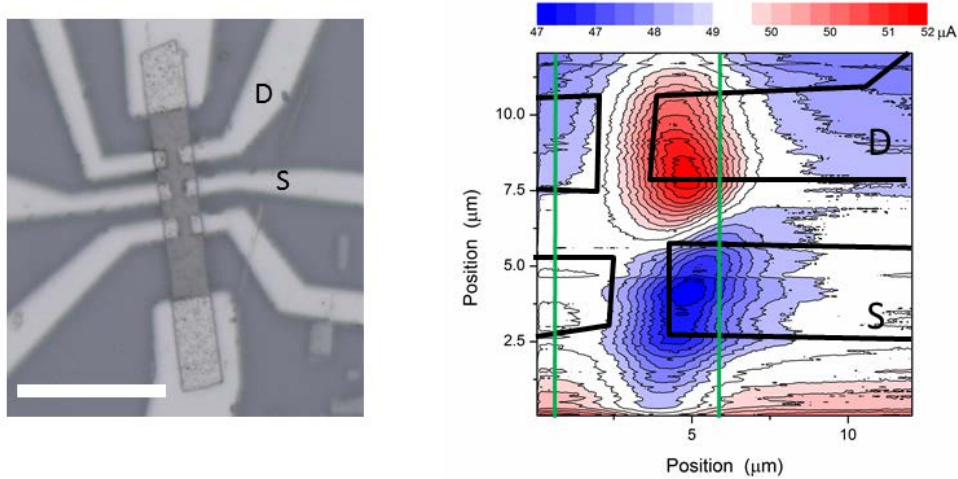

Figure S3: Optical image of a device (Dev SI-3) and the resulting SPCM map.
Scale bar in the optical image is 20 microns. The electrodes between which the photocurrent is measured are indicated with the labels "S" for source and "D" for drain.

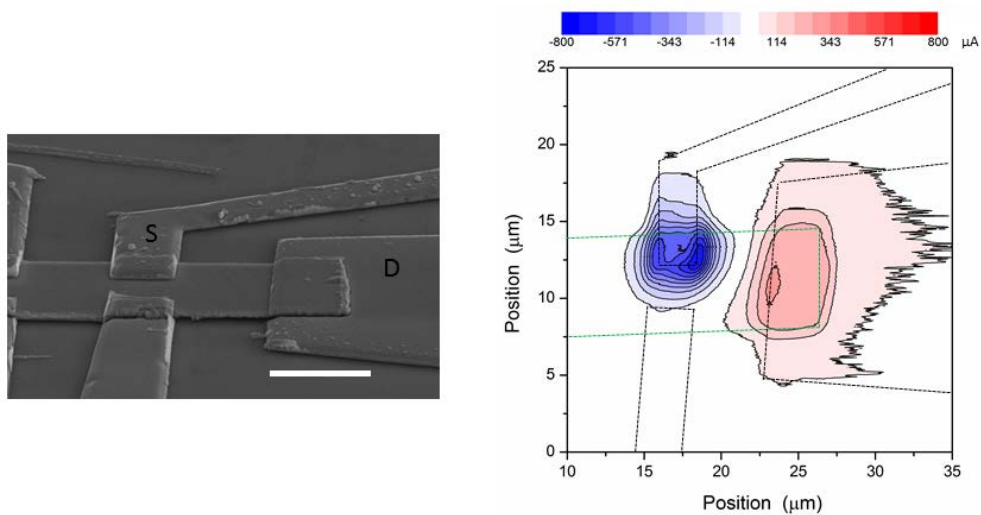

Figure S4: SEM image of a device (Dev SI-4) and the resulting SPCM map.
Scale bar in the SEM image is 5 microns. The electrodes between which the photocurrent is measured are indicated with the labels "S" for source and "D" for drain.

| Device | Responsivity (V/W) | Resistance (Ohm) | Noise Spectrum ($nV/Hz^{1/2}$) | NEP ($pW/Hz^{1/2}$) |
|---|---|---|---|---|
| SI-1 | 17.5 | 665 | 3.3 | 189 |
| SI-2 | 2.22 | 1000 | 4 | 1800 |
| SI-3 | 4 | 200 | 1.8 | 450 |
| SI-4 | 0.85 | 1333 | 4.7 | 5500 |
| Fig. 2 main text | 1.8 | 400 | 2.2 | 1200 |
| Fig. 3 main text | 74 | 569 | 3.1 | 42 |

Table S1: Comparison of the main optoelectronic device properties and performance for several different devices.

2. *Additional examples of device reactivity*

Figure S5 shows SEM images taken on different devices after they had been kept in ambient for several months.

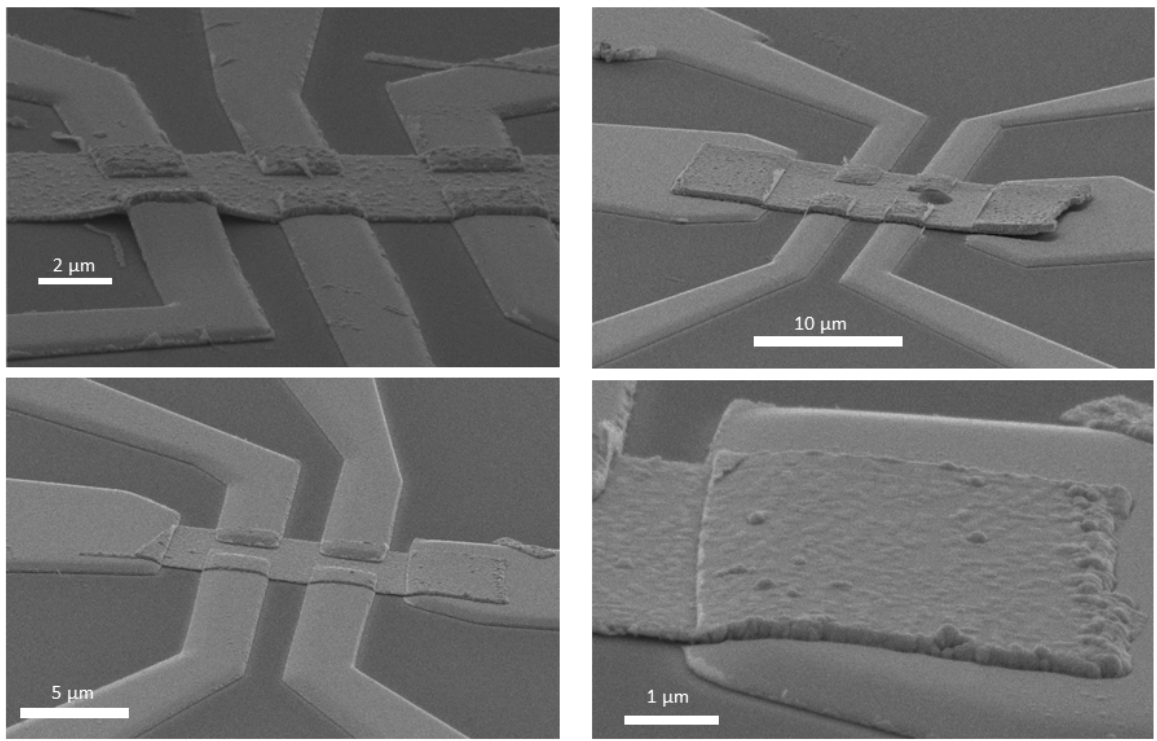

Figure S5: Additional SEM images showing reactivity in different devices.

*3. Response time*

To extract the rise time, we focus the laser on one electrode and measure the time for the signal to rise to 90% of the maximum value. The measurement is repeated with different values of the pre-amplifier response time. An example measurement for a non-suspended device is shown in Fig. S6. For pre-amp integrations times larger than 10 ms the photocurrent rise time is determined by the pre-amplifier. As the pre-amp integration time is reduced below 10 ms, the photocurrent response time saturates to ~5 ms, which is the intrinsic device response time.

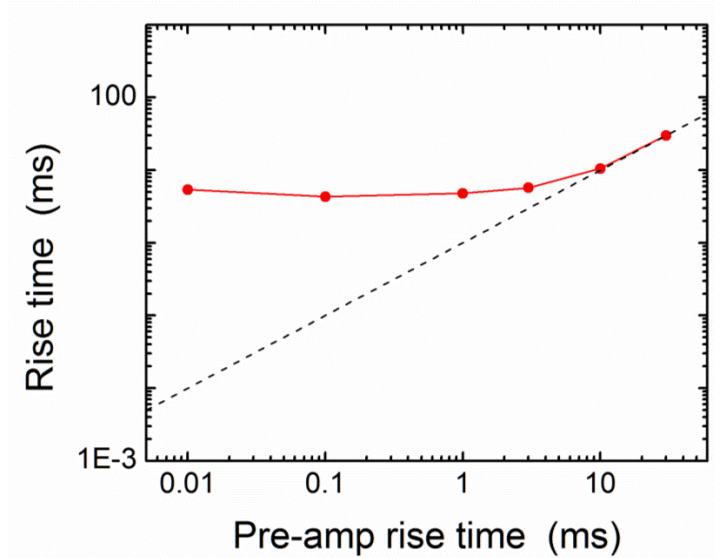

Figure S6: Photocurrent rise time as a function of the pre-amplifier integration time. Here the laser is focused on one electrode.

*4. Transmission Electron Microscopy Analysis*

TEM cross-section specimens were prepared using using standard site-specific focused ion beam (FIB) liftout techniques. Specifically, we employed a FEI Helios Nanolab 660 system. The region of interest was protected by depositing a thin layer of e-beam C, then a layer of i-beam Pt, and finally a layer of i-Beam C. Final polishing of the specimen was conducted using a 5 keV Ga beam.

Observations were made using a probe-corrected FEI Titan instrument equipped with a 4-SDD SuperX energy dispersive x-ray spectroscopy (EDS) detector and operated at 200 keV. The EDS spectrum images were analysed using the Esprit microanalysis software package (version 2.1) (Bruker Nano GmbH, Berlin). For assessing the local compositions, we employed the theoretical Cliff-Lorimer sensitivity factors and x-ray line-shapes provided in the Esprit software.

Figure S7 shows the positions of several regions of interest (ROI) from which we measured compositions from the EDS spectrum images, as tabulated below. In the regions near the bottom of the platelet, the Zr and Te compositions are close to the expected values of 16.7 at. % Zr and 83.3 at. % Te, while the Pd is essentially not detected. In contrast, just below the Pd contact, the Pd compositions are much higher. It is noteworthy that the Zr/Te ratio is near to 0.20, suggesting the Pd may be being incorporated interstitially, with the ZrTe$_5$ lattice remaining. Two examples of EDS spectra extracted from these ROIs are also plotted to illustrate the difference in Pd signal from regions of ZrTe$_5$ platelet taken far from the Pd contact (Region 1) and just below the Pd contact (Region 3).

|  | Region 1 (bottom of ZrTe$_5$) | Region 2 (bottom of ZrTe$_5$) | Region 3 (below Pd contact) | Region 4 (below Pd contact) |
|---|---|---|---|---|
| Zr (at %) | 15.7 ±1.3 | 17.5 ± 1.4 | 14.1 ± 1.3 | 14.2 ± 1.3 |
| Te (at %) | 84.1 ± 8.9 | 82.2 ± 8.8 | 67.1 ± 7.5 | 57.4 ± 6.6 |
| Pd (at %) | 0.2 ± 0.1 | 0.3 ± 0.1 | 18.8 ± 1.9 | 28.4 ± 2.9 |
| Zr/Te ratio | 0.19 | 0.21 | 0.21 | 0.25 |

Table S1. Atomic concentrations at different positions. Uncertainties are quoted as absolute error in the atomic percent at the 1σ level. Additional x-ray signals associated with spectral artifacts due to stray x-rays (Cu, Si, Al) and surface contamination (C, Ga) were included for the spectrum deconvolution, but were not included in the compositional analysis.

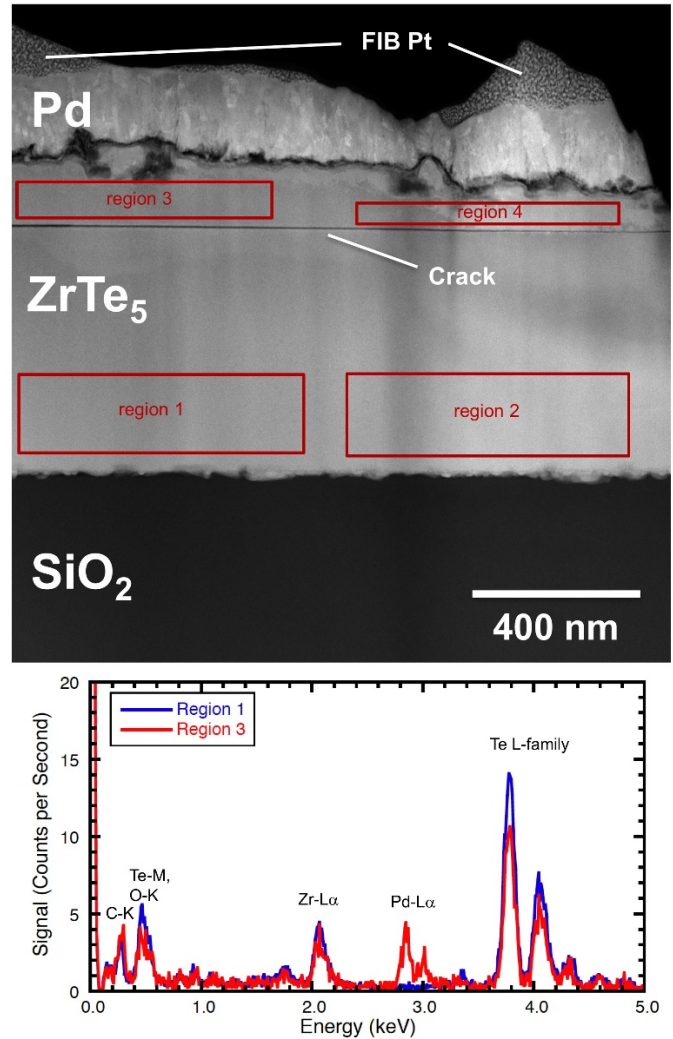

Figure S7. Image shows EDS analysis regions indicated in the table and examples of EDS spectra far from and near to the Pd contact.